\documentclass[%
 aip,jcp,
 amsmath,amssymb,
 preprint,%
]{revtex4-1}

\usepackage{graphicx}
\usepackage{float}
\usepackage{bm}
\usepackage{color}
\usepackage{mathtools}
\usepackage{siunitx}
\usepackage{amsmath}
\usepackage{natbib}
\usepackage{algorithm}
\usepackage{algpseudocode}

\begin{document}

\preprint{AIP/123-QED}

\title{Constant-complexity Stochastic Simulation Algorithm\\with Optimal Binning}%

\author{Kevin R. Sanft}
\email{kevin@kevinsanft.com}
\affiliation{ 
School of Mathematics, University of Minnesota, Minneapolis, MN 55455, USA
}%

\author{Hans G. Othmer}
\email{othmer@math.umn.edu}
\affiliation{ 
School of Mathematics, University of Minnesota, Minneapolis, MN 55455, USA
}%
\affiliation{%
Digital Technology Center, University of Minnesota, Minneapolis, MN 55455, USA
}%

\date{\today}

\begin{abstract}

At the cellular scale, biochemical processes are governed by random interactions between reactant molecules with small copy counts, leading to behavior that is inherently stochastic. Such systems are often modeled as continuous-time Markov jump processes that can be described by the Chemical Master Equation. Gillespie's Stochastic Simulation Algorithm (SSA) generates exact trajectories of these systems. The amount of computational work required for each step of the original SSA is proportional to the number of reaction channels, leading to computational complexity that scales linearly as the problem size increases. The original SSA is therefore inefficient for large problems, which has prompted the development of several alternative formulations with improved scaling properties. We describe an exact SSA that uses a table data structure with event time binning to achieve constant computational complexity. Optimal algorithm parameters and binning strategies are discussed. We compare the computational efficiency of the algorithm to existing methods and demonstrate excellent scaling for large problems. This method is well suited for generating exact trajectories of large models that can be described by the Reaction-Diffusion Master Equation arising from spatially discretized reaction-diffusion processes.

\end{abstract}

\maketitle

\section{\label{sec:intro}Introduction}

Traditional differential equation models of chemical systems work well when the interacting molecules are present in high concentrations. However, at cellular scales, biochemical reactions occur due to the random interactions of reactant molecules present in small numbers. These processes display behavior that cannot be captured by deterministic approaches. Under certain simplifying assumptions, including spatial homogeneity, these systems can be modeled as continuous-time Markov jump processes. The \textit{Chemical Master Equation} (CME) describes the evolution of the probability of the system being in any state at time $t$ \cite{mcquarrie1967, gillespie_cme}. The invariant distribution and the evolution can be found for certain classes of reaction networks \cite{jahnke2007,lee2012,gadgil2005}. However, the CME is too high dimensional to solve for many realistic models. Gillespie's Stochastic Simulation Algorithm (SSA) generates exact trajectories from the distribution described by the CME \cite{Gillespie76,Gillespie77}. Typically, an ensemble of trajectories is run to estimate the probability distribution.

Consider a well-mixed (spatially homogeneous) biochemical system of $S$ different chemical species with associated populations $N(t) = (n_1(t), n_2(t), ..., n_S(t))$. The population is a random variable that changes via $M$ elementary reaction channels $\left\{R_1, R_2, ...,R_M \right\}$. Each reaction channel $R_j$ has an associated \textit{stoichiometry vector} $\nu_j$ that describes how the population $N$ changes when reaction $R_j$ ``fires". The SSA (and CME) are derived by assuming that each reaction channel is a non-homogeneous Poisson process. The stochastic rate or intensity of reaction $R_j$ is determined by a \textit{propensity function}, denoted $a_j$, defined as \cite{Gillespie76,Gillespie77,gillespie2009}:
\begin{align}
\label{eq:propensity_fun}
a_j(N)dt \equiv & \mbox{ probability that reaction } R_j \\
                        & \mbox{ will occur in } [t, t+dt).\nonumber
\end{align}
\noindent The definition in \eqref{eq:propensity_fun} describes a property of an exponential distribution. The time $\tau$ and index $j$ of the next reaction event, given the current state $N$ and time $t$, can therefore be characterized by the joint probability density function
\begin{equation}
\label{eq:joint_density}
p(\tau, j | N, t) = e^{ - \sum_{i=1}^{M} a_i(N) \tau}a_j(N).
\end{equation}
\noindent Equation \eqref{eq:joint_density} combined with the fact that the process is Markovian (a consequence of the well-mixed assumption) leads naturally to the SSA.

We will refer to the (exact) ``SSA" as any simulation algorithm that generates trajectories by producing exact samples of $\tau$ and $j$ from the density in \eqref{eq:joint_density}. Note that when other sources describe ``the SSA" or ``the Gillespie algorithm", they are often implicitly referring to the \textit{direct method} variant of the SSA (see Section \ref{sec:direct_method}) \cite{Gillespie76,Gillespie77}. However, many other SSA variants have been proposed that achieve different performance and algorithm scaling properties by utilizing alternate data structures to sample the density function \eqref{eq:joint_density} \cite{NextRxnMethod,composition_rejection, nested_ssa, mauchEfficient, flavor, ramaswamy2009new, yates2013recycling, anderson_2007, hu_2014}. Many of these methods apply random variate generation techniques described in Devroye \cite{devroye1986} to the SSA. Among the most popular of these alternate formulations of the SSA for large problems is the \textit{Next Reaction Method} (NRM) of Gibson and Bruck \cite{NextRxnMethod}. We discuss the NRM and other formulations in detail in later sections. Many authors have also described \textit{approximate} methods that sacrifice exactness in exchange for computational efficiency gains, including the class of \textit{tau-leaping} algorithms and various hybrid approaches \cite{tau, ssSSA, dynamic_partitioning,haseltine_rawlings2002, salis_kaznessis_hybrid_2005, hy3s, yang_stiffness_detection_2011, rem_tau, ferm_j_sci_comput_2008, implicit_tau, rao_arkin}. In this paper, we restrict our attention to exact methods. 

An important class of problems in which large models arise come from spatially-discretized reaction diffusion processes. The assumption underlying the CME and SSA that the reactant molecules are spatially homogeneous is often not justified in biological settings. One way to relax this assumption is to discretize the system volume into $N_s$ subvolumes and assume that the molecules are well-mixed within each subvolume. Molecules are allowed to move from one subvolume to an adjacent one via diffusive transfer events. This setting can also be described as a Markov jump process and leads to the reaction-diffusion master equation (RDME) \cite{gardiner1976,gillespie2013}. In simulating the RDME, reaction events within each subvolume are simulated with the standard SSA and diffusive transfers are modeled as pseudo-first order ``reactions". The resulting simulation method is algorithmically equivalent to the SSA (see Gillespie \textit{et al.} \cite{gillespie2013} for an overview). However, the state space and the number of transition channels grows quickly as the number of subvolumes increases. The population of every species must be recorded in every subvolume and the number of transition channels includes all diffusive transfer channels plus each reaction channel in the homogeneous system is effectively duplicated once for each subvolume. In a spatial model, $M$ is the total number of reaction and diffusion channels.

In this paper we present an exact SSA variant that is highly efficient for large problems. The method can be viewed as a variation of the NRM that uses a table data structure in which event times are stored in ``bins". By choosing the bin size relative to the average simulation step size, one can determine an upper bound on the average amount of computational work per simulation step, independent of the number of reaction channels. This constant-complexity NRM is best suited for models where the fastest timescale stays relatively constant throughout the simulation. Optimal performance of the algorithm is sensitive to the binning strategy, which we discuss in detail in Section \ref{sec:optimal_binning}.

The remainder of the paper is organized as follows. In the next section, we discuss the standard direct and NRM methods in more detail. In Section \ref{sec:alternate_formulations}, we demonstrate how alternate methods with various scaling properties can be derived by modifying the standard methods. Section \eqref{sec:constant_nrm} presents the constant-complexity NRM algorithm and the optimal binning strategy. The performance and scaling of several methods is presented in Section \ref{sec:numerical_experiments}. Section \ref{sec:conclusions} concludes with a brief summary and discussion.

\section{Standard Methods for Spatially Homogeneous Models}
\label{sec:standard_methods}

The majority of exact SSA methods are based on variations of the two standard implementations: the direct method and the NRM. Here we review these standard methods in more detail.

\subsection{Direct Method}
\label{sec:direct_method}

The \textit{direct method} variant of the SSA is the most popular implementation of the Gillespie algorithm \cite{Gillespie76,Gillespie77}. In the direct method, the step size $\tau$ is calculated by generating an exponentially distributed random number with rate parameter (or intensity) equal to the sum of the propensities. That is, $\tau$ is chosen from the probability density function
\begin{equation}
f(\tau) = \left(\sum_{i=1}^{M} a_i(N)\right) e^{- \sum_{i=1}^{M} a_i(N)\tau},\;\; \tau \ge 0.
\end{equation}
\noindent The index of the next reaction, $j$, is an integer-valued random variable chosen from the probability mass function
\begin{equation}
\label{eq:j_pmf}
f(j) = \frac{a_j}{\sum_{i=1}^{M} a_i(N)}, \;\; j \in \{1, 2, ..., M\}.
\end{equation}
\noindent The direct method is typically implemented by storing the propensities in a one dimensional data structure (e.g. an array in the C programming language). Generating the reaction index $j$ as described in \eqref{eq:j_pmf} can be implemented by choosing a continuous uniform random number $r$ between $0$ and the propensity sum and choosing $j$ such that
\begin{equation}
\label{eq:select_j}
j = \mbox{ smallest integer satisfying } \sum_{i=1}^{j} a_i(N) > r.
\end{equation}
\noindent This ``search" to find the index of the next reaction proceeds by iterating over the elements in the array until condition \eqref{eq:select_j} is satisfied. After the time step size and reaction index are selected, the simulation time $t$ and system state $N$ are updated as $t = t + \tau$ and $N = N + \nu_j$. Changing the population $N$ will generally lead to changes in some propensities. These affected propensities are recalculated by evaluating the propensity functions for the new population $N$ and the new propensity values replace the old values in the propensities array data structure. The propensity sum is adjusted accordingly. The algorithm repeats until a termination condition is satisfied. Typically, an ensemble of many simulation trajectories is computed. To begin each simulation, the system time is reset to zero and the state vector is set to the initial condition and the algorithm is repeated.

An analysis of the scaling properties for a single step of the direct method, or any SSA variant, can be done by considering the computational work required for the two primary tasks of the algorithm: 1) searching for the next reaction index, and 2) updating the reaction generator data structure (see footnote \label{build_footnote} \cite{footnote1}). If the propensities are stored in no particular order in the linear array, the search will take approximately $M/2$ steps on average. If knowledge of the average magnitudes of the propensities is available, then the average search can be shortened from $M/2$ to some smaller fraction of $M$ by utilizing static or dynamic sorting, which can speed up the simulation \cite{odm,sdm}. In both the sorted and unsorted case, the search step has computational cost that scales as $O(M)$. In updating the propensity array data structure, the following operations occur for each propensity that must be updated: the propensity function is evaluated, the propensity value is updated in the propensity array, and the propensity sum is updated. In a typical implementation, these are all $O(1)$ operations as the amount of computational work does not depend on the problem size. The total cost of the update step therefore depends on how many propensities must be updated when a given reaction occurs. The average cost of the update step will be this number of dependencies for each reaction, weighted by the firing frequencies of the reactions. The total cost of each step of the direct method is therefore $O(M)$ for the search for the next reaction plus an update cost $D$, where $D$ is proportional to the average number of propensity updates required per simulation time step. We assume that $D$ is bounded above by a constant independent of $M$, making the direct method an $O(M)$ algorithm overall. The SSA direct method is easy to implement with simple data structures that have low overhead, making it a popular choice that performs well for small problem sizes.

\subsection{Next Reaction Method}

In his original 1976 paper\cite{Gillespie76}, Gillespie also proposed the \textit{first reaction method}, which involves generating the (tentative) next reaction time for every reaction channel and then finding the smallest value to determine the next event that occurs. Like the direct method, this can be implemented using a simple 1D array data structure. The smallest event time is then found by iterating over all the elements. The direct method and first reaction method both generate exact simulation trajectories and both scale as $O(M)$. However, in practice the direct method tends to be more efficient for standard implementations, primarily because the direct method requires about $M/2$ operations per step whereas the first reaction method requires $M$ operations to find the smallest element.

The NRM is a variation of the first reaction method that uses a binary min-heap to store the next reaction times \cite{NextRxnMethod}. A binary min-heap is a complete binary tree in which each node contains a value that is less than (or equal to) its children. The search for the next reaction is therefore a constant-complexity, or $O(1)$, operation because the next reaction corresponds to the smallest reaction time, which is always located at the top of the heap. However, updating the binary min-heap data structure when a propensity changes is computationally expensive. For each affected propensity, the propensity function has to be evaluated, the reaction time has to be recomputed based on this new propensity, and the binary min-heap must be updated to maintain the heap property. Maintaining the heap structure generally requires $O(log_2(M))$ operations for each affected event time, implying that the total computational cost of the update step is $DO(log_2(M)) = O(log_2(M))$, where $D$ is again the average number of propensity updates required per step \cite{NextRxnMethod}. In practice, the actual update cost will depend on many factors, including the extent to which the fastest reaction channels are coupled and whether or not the random numbers are ``recycled" \cite{NextRxnMethod, anderson_2007}. The interested reader should consult Gibson and Bruck \cite{NextRxnMethod} for a detailed analysis. The $O(log_2(M))$ scaling makes the NRM superior to the direct method for large problems.

\section{Alternative Formulations}
\label{sec:alternate_formulations}

There are two main techniques for improving the scaling properties of an exact SSA implementation. The first is to use different data structures, as in the binary tree utilized for the NRM min-heap implementation. The other technique is to split the search problem into smaller subproblems by partitioning the reaction channels into subsets. The two techniques are complementary, and in some cases conceptually equivalent. By utilizing different combinations of data structures and partitioning schemes, it is possible to define an infinite number of alternate SSA formulations with varying performance and scaling properties.

To fix ideas, we consider a simple variation of the direct method. Instead of storing all $M$ propensities in a single array, one could partition the reaction channels into two subsets of size $M/2$ and store them using two arrays. If we label the propensity sums of the two subsets $a_{S_1}$ and $a_{S_2}$, respectively, then the probability that the next reaction will come from subset $i$ is $a_{S_i}/(a_{S_1}+a_{S_2})$. Conditioned on the next reaction coming from subset $i$, the probability that the next reaction is $R_j$ will be $a_j/a_{S_i}$, for all $j$ in subset $i$. These statistical facts follow from the properties of exponential random variables and lead naturally to a simulation algorithm. We first select the subset $i$ from which the next reaction will occur. To do this, we choose a continuous uniform random number $r_1$ between 0 and the propensity sum and then chose $i$ such that
\begin{equation}
\label{eq:select_subset}
i = \mbox{ smallest integer satisfying } \sum_{k=1}^{i} a_{S_k}(N) > r_1.
\end{equation}
The similarity to \eqref{eq:select_j} is apparent as we are essentially performing the direct method's linear search but applied to subset propensity sums instead of propensities. We have written \eqref{eq:select_subset} in a generic way to accommodate more than two subsets, which we will consider in the next subsection. Once we have determined the subset $i$ from which the next reaction will occur, we can use the direct method's linear search from \eqref{eq:select_j} again to select the reaction $j$ from within subset $i$, but choosing the uniform random number between 0 and $a_{S_i}$ and iterating over only the elements in the array corresponding to subset $i$.

The resulting algorithm can be viewed as a ``2D" version of the direct method. The algorithm requires, on average, a search of depth 1.5 to choose the subset, assuming the reaction channels were partitioned in no particular order. The search within the subset will then require an average search depth of $M/4$ to select the particular reaction within the chosen subset. By partitioning the reaction channels into two subsets, we have derived an algorithm that is statistically equivalent to the direct method but with different scaling properties. The new method has slightly more overhead than the original direct method because we must update the subset propensity sums at each step. It will therefore be less efficient than the original direct method for small problems. However, the $M/4$ scaling will outperform the $M/2$ scaling of the direct method for larger problems. The algorithmic complexity of this variation is still $O(M)$, so the NRM will also outperform this method for sufficiently large problems. Below we show how more sophisticated partitioning and data structures can be used to achieve improved scaling properties.

\subsection{Alternate Direct Formulations}
\label{subsec:alternate_direct}

One can expand on the idea above to create other direct method variants. If each of the two subsets is then split into two more subsets, a similar procedure can be applied to derive a method with $M/8$ scaling. If this processes is repeated recursively, the resulting partitioning of the reaction channels can be viewed as a binary tree \cite{NextRxnMethod,mauchEfficient,hu_2014}. This \textit{logarithmic direct method} has $O(log_2(M))$ cost for the search and update steps. If instead of partitioning into two subsets, the reaction channels were partitioned into $\sqrt{M}$ subsets each containing $\sqrt{M}$ channels, the search for the subset and the search within the subset are both $O(\sqrt{M})$ operations, leading to $O(\sqrt{M})$ algorithmic complexity. This is the theoretically optimal ``2D search" formulation. Similarly, one can define a three-dimensional data structure, leading to a ``3D search" which scales as $O(\sqrt[3]{M})$, and so on \cite{mauchEfficient,hu_2014}.

Slepoy \textit{et al.} \cite{composition_rejection} proposed a constant-complexity (O(1)) algorithm that uses a clever partitioning strategy combined with rejection sampling. First, the reaction channels are partitioned based on the magnitude of their propensities with partition boundaries corresponding to propensities equal to powers of two. That is, if we number the partitions using (positive and negative) integers, partition $i$ contains all of the reaction channels with propensities in the range $[2^i, 2^{i+1})$. The particular partition $g$ from which the next reaction will occur is chosen exactly via a linear search. The average search depth to select the subset will depend on the range between the largest and smallest nonzero propensity value. This search is assumed to have a small average search depth, independent of $M$. Once the subset $g$ from which the next reaction will occur is determined, a rejection sampling technique is employed to select the particular channel within the subset. An integer random number $r_1$ is chosen between $[1, M_g]$, where $M_g$ is the number of reaction channels in the chosen subset. This tentatively selects the reaction channel $j$ corresponding to the $r_{1}^{th}$ channel in the subset. Then a continuous random number $r_2$ is chosen between $[0, 2^{g+1})$. If $r_2 < a_j(x)$, reaction channel $j$ is accepted as the next event, otherwise it is rejected and new random numbers $r_1$ and $r_2$ are chosen. The process is repeated until a reaction is accepted. This procedure selects the reaction according to the exact probability density function. Since the subsets have been engineered such that all propensities in subset $g$ are within the range $[2^g, 2^{g+1})$, on average less than half of the selected reactions will be rejected. Therefore, it takes fewer than two samples on average to select the next reaction, independent of the number of reaction channels. Updating the data structure for each affected propensity is also an amortized constant-complexity operation that requires, in the worst case, removing an element from one partition and inserting it into a different partition, which occurs whenever the change in propensity crosses a power of two boundary. The search and update steps are both constant-complexity, leading to $O(1)$ algorithmic complexity independent of $M$.

\subsection{Next Subvolume Method}

The preceding methods are general formulations that can be applied to any model that can be described by the CME. In this subsection we describe the \textit{next subvolume method} (NSM, not to be confused with the NRM), which is formulated specifically for simulating processes described by the RDME. The NSM is a variation of a 3D search method \cite{elf2004spontaneous} that partitions the channels (reaction and diffusion) based on the spatial structure of the model. The NSM has several desirable properties that make it efficient, and hence popular, for simulating spatial models. The NSM first partitions on subvolumes and uses the NRM (implemented using a binary min-heap) to select the subvolume in which the next event occurs. Within each subvolume, the diffusion channels and reaction channels are stored using two arrays, similar to the two-partition direct method scheme from the beginning of this section. To choose the particular event within the subvolume, the NSM first does a linear search to determine whether the next reaction is diffusion or a reaction, then chooses the particular event channel via a linear search within that partition. Among the favorable properties of the NSM is that organizing the events by subvolume tends to keep the updates local in memory. Partitioning by subvolume ensures that an event affects at most two subvolumes (which occurs when the event is a diffusive transfer), therefore, at most two values in the binary min-heap need to be updated on each step of the algorithm. The NSM scales as $O(log_2(N_s))$, where $N_s$ is the number of subvolumes. In practice, the NSM performs well on spatial models spanning a wide range of problem sizes.

Recently, Hu \textit{et al.} presented a method in which the NSM search is effectively reversed \cite{hu_2014}. First, the type of event is selected, then the subvolume is chosen using a binary, 2D, or 3D search, leading to algorithmic complexity that is $O(log_2(N_s)$, $O(\sqrt{N_s})$, or $O(\sqrt[3]{N_s})$, respectively.

\section{Constant-complexity Next Reaction Method}
\label{sec:constant_nrm}

The NRM was derived by taking the basic idea of the first reaction method and utilizing a different data structure to locate the smallest event time. The abstract data type for this situation is known as a \textit{priority queue} and in principle any correct priority queue implementation can be used. Here we present an implementation of the next reaction method that uses a table data structure comprised of ``bins" to store the event times for the priority queue. If the propensity sum, and therefore the expected simulation step size, can be bounded above and below, then the average number of operations to select the next reaction and update the priority queue will be bounded above, independent of M, resulting in a constant-complexity NRM.

To implement a constant-complexity NRM, we partition the total simulation time interval, from the initial time $t=0$ to final time $t=T_f$, into a table of $K$ bins of width $W$. For now, we will let $WK = T_f$. Bin $B_i$ will contain all event times in the range $[iW, (i+1)W)$. We generate putative event times for all reaction channels as in the original NRM, but insert them into the appropriate bin in the table instead of in a binary min-heap. Events that fall outside of the table range are not stored in the table. Values within a bin are stored in a 1D array (though alternative data structures could be used). To select the next reaction, we must find the smallest element. Therefore, we must find the first nonempty bin (i.e. the smallest $i$ such that $B_i$ is not empty). To locate the first non-empty bin, we begin by considering the bin $i$ from which the previous event was found. If that bin is empty, we repeatedly increment $i$ and check the $i^{th}$ bin until a nonempty bin is found. We then iterate over the elements within that bin to locate the smallest value. 
The update step of the algorithm requires computing the new propensity value and event time for each affected propensity. In the worst case, the new event time will cause the event to move from one bin to another, which is an $O(1)$ operation. Therefore, the update step will be an $O(D)$ operation, where again $D$ is the average number of propensity updates required per simulation step. We assume that $D$ is bounded above independent of $M$. Therefore, if the propensity sum is bounded above and below independent of $M$, the overall complexity of the algorithm is $O(1)$. In the next section we consider the optimal bin width to minimize the cost to select the next reaction.

\begin{figure}
	\centering
	\includegraphics[scale=0.3]{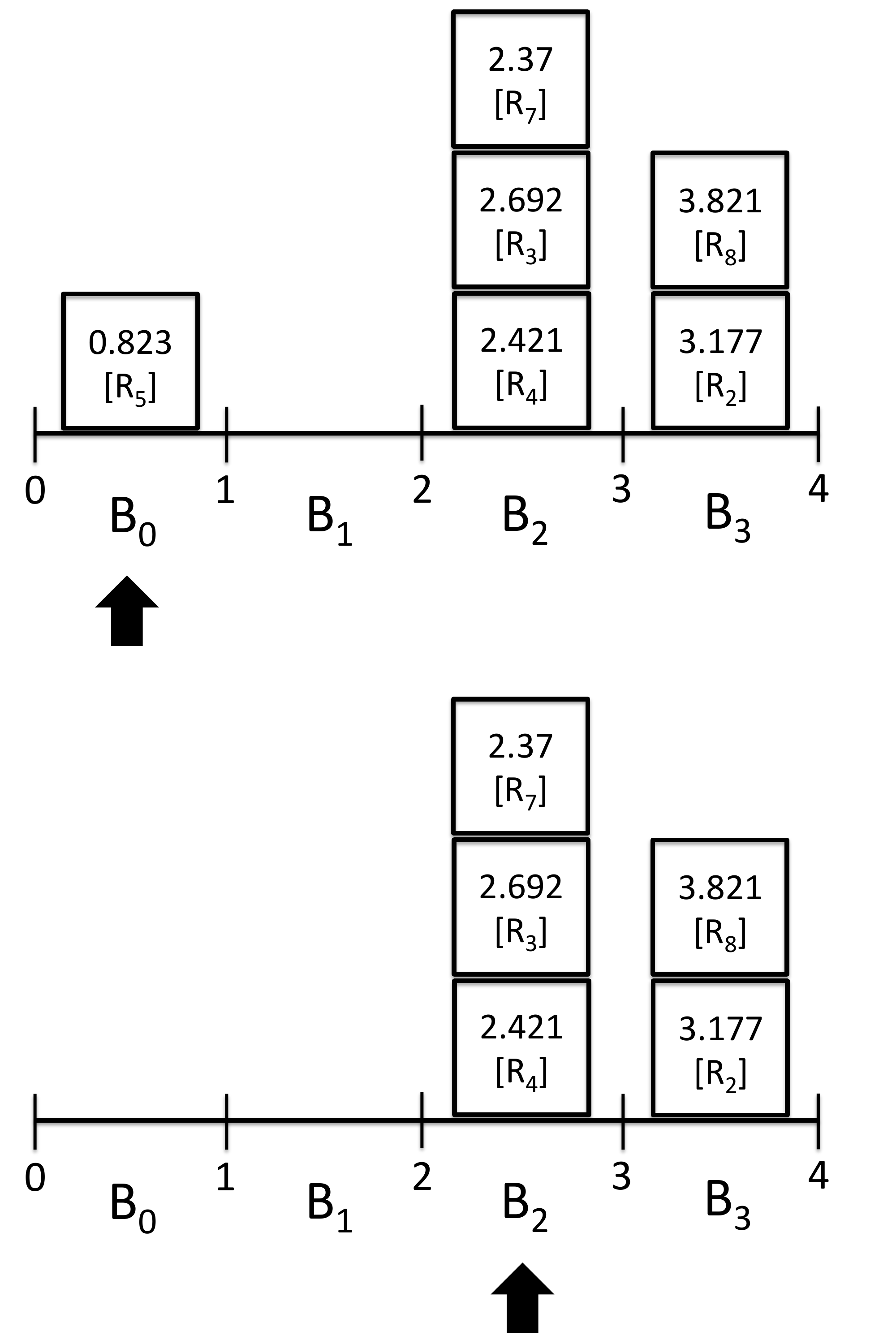}
	\caption{Priority queue using binning. The next reaction in a simulation is the event with the smallest event time. Locating the smallest event time requires first locating the smallest non-empty bin, then finding the smallest element within that bin. Elements in the bins contain the event time and the event (reaction) index. Bin boundaries are integers here for simplicity. Elements within a bin are unsorted. Elements with zero propensity or with an event time beyond the last bin are not stored in the queue. In the top figure, the first bin, $B_0$, contains the smallest element as indicated by the arrow. After that event occurs, the search for the next nonempty bin begins at the current bin and, if empty, the bin pointer is incremented until the next nonempty bin is located, as indicated by the arrow under bin $B_2$ in the bottom figure. Here the next event is ``$R_7$". In a real simulation, the update step may cause elements to move from one bin to another.}
	\label{fig:fig1}
\end{figure}

\subsection{Optimal Bin Width}
\label{sec:optimal_binning}

Storing the elements within a bin using a 1D array is similar to the chaining approach (sometimes referred to as ``closed addressing") to collision resolution in a hash table. A hash table is a data structure in which values are mapped to bins using a \emph{hash function}. Since there are generally more possible values than total bins, collisions, where multiple values are mapped to the same bin, are possible. Chaining is a collision resolution strategy where multiple values that map to the same bin are stored in a linked list. A well-designed hash table typically has amortized constant time insertion and retrieval. (Interested readers should consult an introductory computer science data structures textbook.) An important difference between our data structure and a hash table implementation is that hash tables are typically designed to have a particular \textit{load factor}. The load factor is defined as the number of elements stored in the table divided by the number of bins. For a hash table with a good hash function, a low load factor ensures that each bin will contain a small number of elements, on average. However, we are effectively using the reaction event times as a hash function. Whereas a good hash function distributes elements uniformly amongst the bins, the reaction times are not uniformly distributed, they are \textit{exponentially} distributed. Targeting a particular load factor could lead to good or bad performance depending on the distribution of the propensities, because the key to efficiency is choosing the appropriate bin width relative to the mean simulation step size. If the bin width is too small, there will be many empty bins and if the bin width is too large there will be many elements within each bin.

In considering the search cost and optimal bin width, it is helpful to consider two extreme cases. For the first case, suppose one reaction channel is ``fast" and the rest are slow. For the second case, suppose all reaction channels have equal rates (propensities). For simplicity of analysis we can rescale time so that the propensity sum equals one and we assume $T_f \gg 1$. Then in the first case, if the propensity of the fast channel is much larger than the sum of the slow propensities (i.e. the fast propensity is approximately equal to one), we can choose the bin width to be large on the scale of the fast reaction but small on the scale of the slow reactions. For example, choosing $W \approx 6.64$ (corresponding to the $99^{th}$ percentile for a unit rate exponential distribution) ensures that the fast reaction will initially be in the first bin with approximately $99\%$ probability. By assumption, there is a small probability that any of the slow reactions will appear in the first bin. Upon executing the simulation, the fast reaction will repeatedly appear in the first bin and be selected during the next step, until it eventually appears in the second bin (with probability $< 1\%$ of landing beyond the second bin). If it takes on average roughly 6.6 steps before the fast reaction appears in the second bin, then the average search depth to locate the first nonempty bin is about $1 + \frac{1}{6.6} \approx 1.15$ and the average search depth within a bin is approximately one. The ``total search depth" will be approximately $2.15$. The slow reactions contribute a relatively small additional cost in this scenario. If, however, the slow reactions are not negligible, then the fast reaction plays a less important role in the search cost and the situation can be viewed as similar to the second case, which we consider next.

Here we suppose that all reaction channels have equal propensities and the propensity sum equals one. In this case, the number of elements that will initially be placed in the first bin will be approximately Poisson distributed with mean $W$. As the simulation progresses, elements will be removed from the first bin until it is emptied and the simulation will move on to the second bin where the process repeats. If the number of events per bin is Poisson distributed with rate $W$, the average search depth to locate the first nonempty bin is $1/W + 1$ and the average search depth within a bin is $W/2 + 1$. The total search depth is minimized when $W = \sqrt{2}$, leading to a total search depth of $2 + \sqrt{2} \approx 3.41$. That is, it takes an average of about $3.41$ operations to select the next reaction, independent of the size of the model. If the propensity sum does not equal one, this minimum total search depth will be achieved with a bin width of $W = \sqrt{2} \sum_{i=1}^{M} a_i(x)$. 

\begin{figure}[h]
\begin{center}
\includegraphics[scale=0.5]{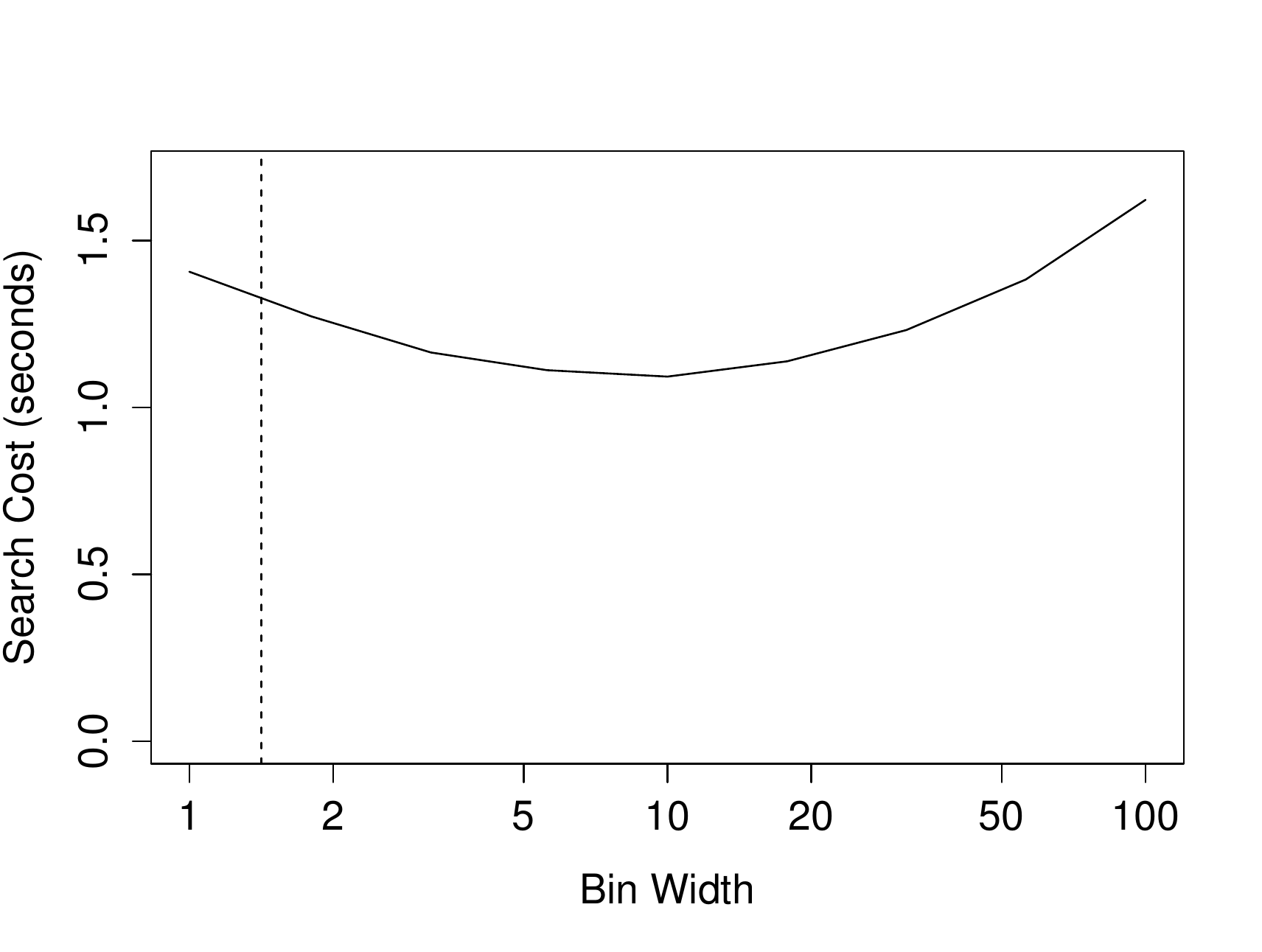}
\end{center}
\caption{Search cost for various bin widths. In this example, $M = 10^7$ and every propensity $= 10^{-7}$ (the propensity sum equals one). The theoretical optimal bin width for minimizing the total search depth corresponding to $W = \sqrt{2}$ is shown as a dashed vertical line. In practice, the true optimal bin width is larger than $W = \sqrt{2}$. A bin width of 16 times the mean simulation step size performs well over a wide range of problem sizes. Search cost is measured in seconds for $10^7$ simulation steps.}
\label{fig:fig2}
\end{figure}

The theoretical optimal relative bin width $W = \sqrt{2}$ does not minimize the search cost in an actual implementation. Figure \ref{fig:fig2} shows that the search cost is minimized at a bin width much larger than $\sqrt{2}$. One reason for this is that accessing consecutive items within a bin is generally faster than traversing between bins because items within a bin are stored in contiguous blocks of memory. In our experience, a bin width of approximately 16 times the mean simulation step size performs well across a wide range of problem sizes. Widths between 8 and 32 times the step size perform well, making the choice of 16 robust to modest changes in the propensity sum. However, it is possible that the optimal values may vary slightly depending on the system architecture. More importantly, if the propensity sum varies over many orders of magnitude during a simulation, a static bin width may be far from optimal during portions of the simulation.

\subsection{Optimal Number of Bins and Dynamic Rebuilding}

\begin{figure}[h]
\begin{center}
\includegraphics[scale=0.4]{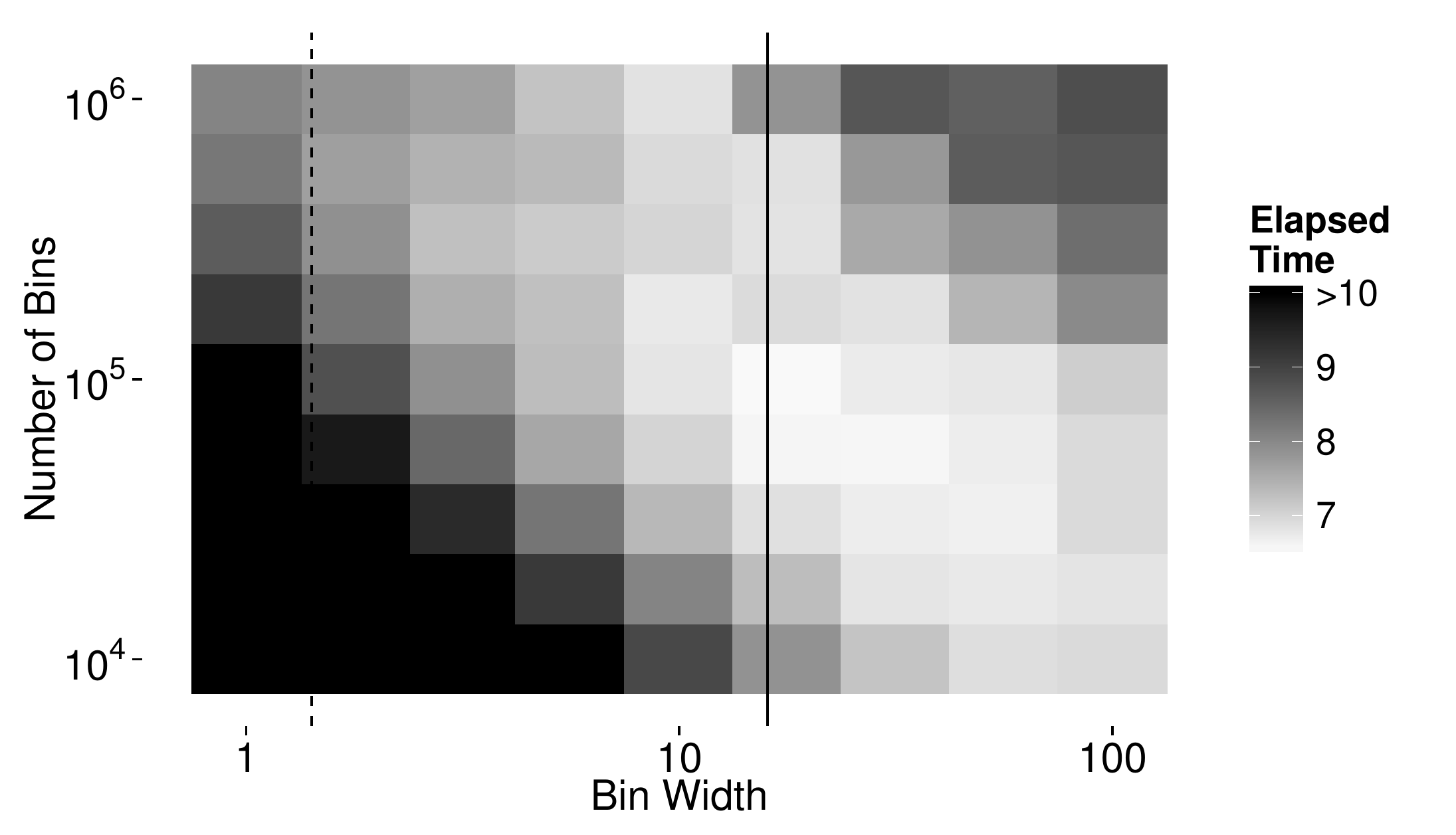}
\end{center}
\caption{Simulation time for various bin widths and number of bins. In this example, $M = 10^7$ and every propensity $= 10^{-7}$ (the propensity sum equals one). The dashed line corresponds to the theoretical optimal bin width $W = \sqrt{2}$ that minimizes the total search depth. The solid line corresponds to $W = 16$, which is the target bin width used in practice. The algorithm performs well over a fairly wide range of bin widths and number of bins. As the problem size increases, the optimal number of bins increases roughly proportional to $\sqrt{M}$.}
\label{fig:fig3}
\end{figure}

It is not necessary to choose the number of bins $K$ such that $WK \ge T_f$, where again $T_f$ is the simulation end time. Clearly, choosing $K$ such that $W(K-1) > T_f$ means that the table is larger than necessary, which is inefficient as larger memory use leads to slower simulations. However, it is less obvious that a choice of $K$ such that $WK < T_f$ can lead to improved performance. When $WK < T_f$, the table data structure must be rebuilt when the simulation time exceeds $WK$. A tradeoff exists between choosing $K$ large, which is less efficient because it uses more memory, and choosing $K$ small, which requires more frequent rebuilding of the table data structure. The propensity and reaction time ``update step" also benefits slightly from a smaller table because fewer reaction channels will be stored in the table leading to fewer operations required to move reactions from one bin to another. Figure \ref{fig:fig3} shows the elapsed time to execute simulations for varying bin widths and numbers of bins.

As the problem size increases, the optimal number of bins gets larger due to the increased rebuild cost. We have found that the optimal number of bins scales roughly proportional to the square root of the number of reaction channels. In practice, choosing $K = 20\sqrt{M}$ leads to good performance across a wide range of problem sizes, though the optimal value may vary across different system architectures. In the case where many of the reaction channels have zero propensity, it is more efficient to use the average number of nonzero propensity channels instead of $M$ in computing the number of bins. To facilitate rebuilding the table, we record the number of steps since the last rebuild. This allows for the bin width to be chosen adaptively based on the simulation step sizes used most recently. This adaptive bin sizing strategy partially mitigates the problem of suboptimal bin widths that may arise due to changing propensity sums. Overall, the constant-complexity NRM algorithm with a fixed target relative bin width and dynamic rebuilding strategy exhibits excellent efficiency across a wide range of problem sizes as demonstrated in the next section.

\section{Numerical Experiments}
\label{sec:numerical_experiments}

In this section we demonstrate the performance and scaling of the constant-complexity NRM (``NRM (constant)") relative to other popular methods. Among the other methods considered are the constant-complexity direct method (``Direct (constant)" composition-rejection algorithm) of Slepoy \emph{et al.} \cite{composition_rejection}, the original NRM (``NRM (binary)") of Gibson and Bruck \cite{NextRxnMethod}, and the NSM of Elf and Ehrenberg \cite{elf2004spontaneous}. The algorithms were implemented in C++, using code from StochKit2 \cite{StochKit2} and Cain \cite{cain} where possible. Pseudocode that outlines the constant-complexity NRM is given in Appendix A. All timing experiments were conducted on a Macbook Pro laptop with a 2.4 GHz Core i5 processor and 8 GB of memory.

\subsection{Reaction Generator Test}

Most exact SSA variants can be viewed as either a Direct Method or NRM implementation with varying data structures used to select the next reaction. The performance of the reaction generator data structure is the primary determinant of the overall algorithm performance. In this section we test the efficiency of several reaction generator data structures, independent of the rest of the solver code, by simulating the ``arrivals" of a network of $M$ Poisson processes.

\begin{figure}[h]
\begin{center}
\includegraphics[scale=0.4]{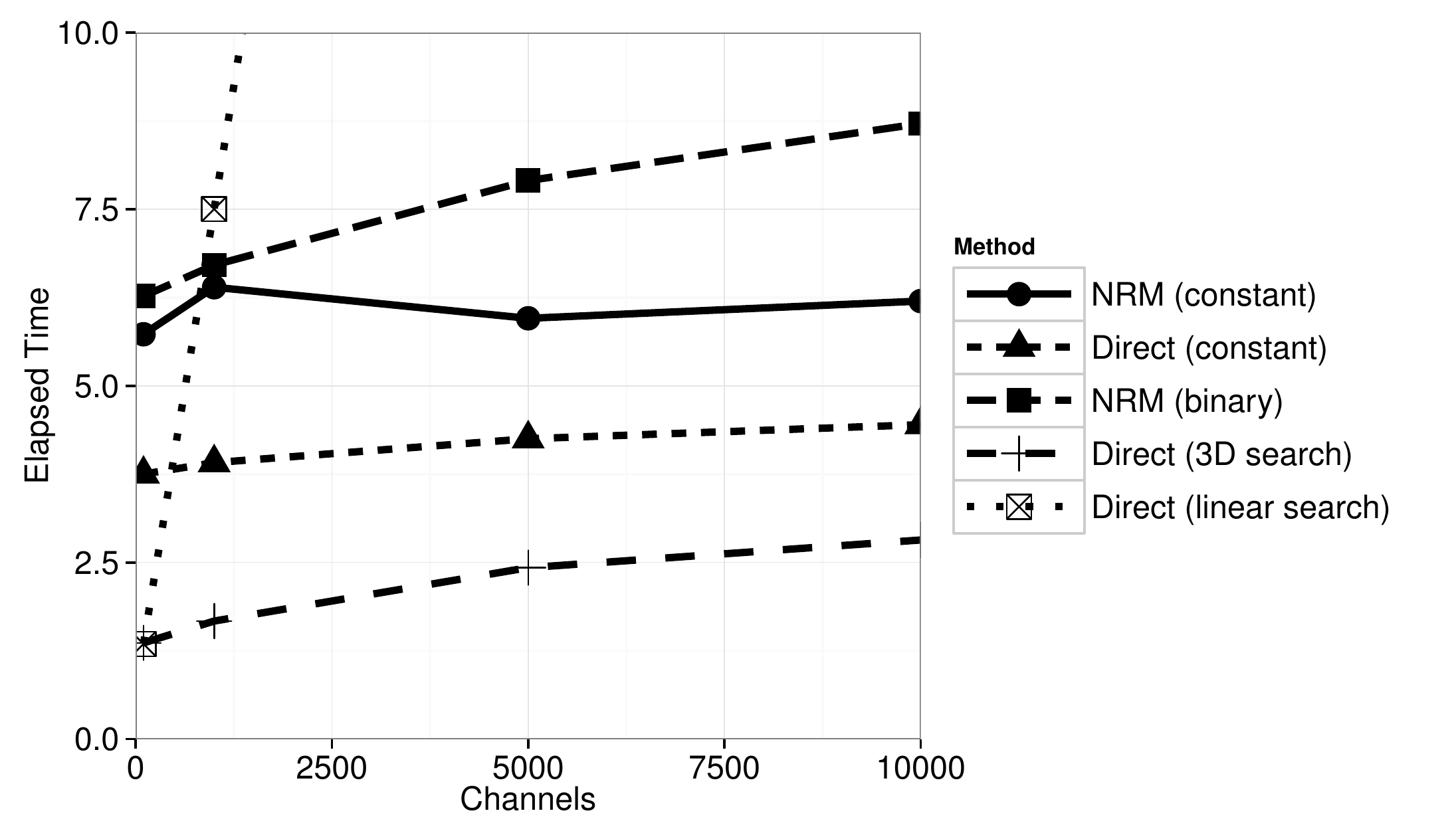}
\end{center}
\caption{Scaling on small problems. Elapsed time in seconds to generate the reaction index and update the data structure $10^7$ times for various reaction generators. Each reaction channel has a unit rate propensity and a random network in which 10 updates are required was generated. For extremely small models, where $M < 100$ or so, the original direct method with linear search performs best. As the problem size increases, the direct method with a 3D search is optimal. Not shown is the direct method with 2D search, which slightly outperforms 3D search when $M < 5000$. The constant-complexity NRM performance exhibits some fluctuations because the implementation was not optimized for small problems.}
\label{fig:fig4}
\end{figure}

\begin{figure}[h]
\begin{center}
\includegraphics[scale=0.4]{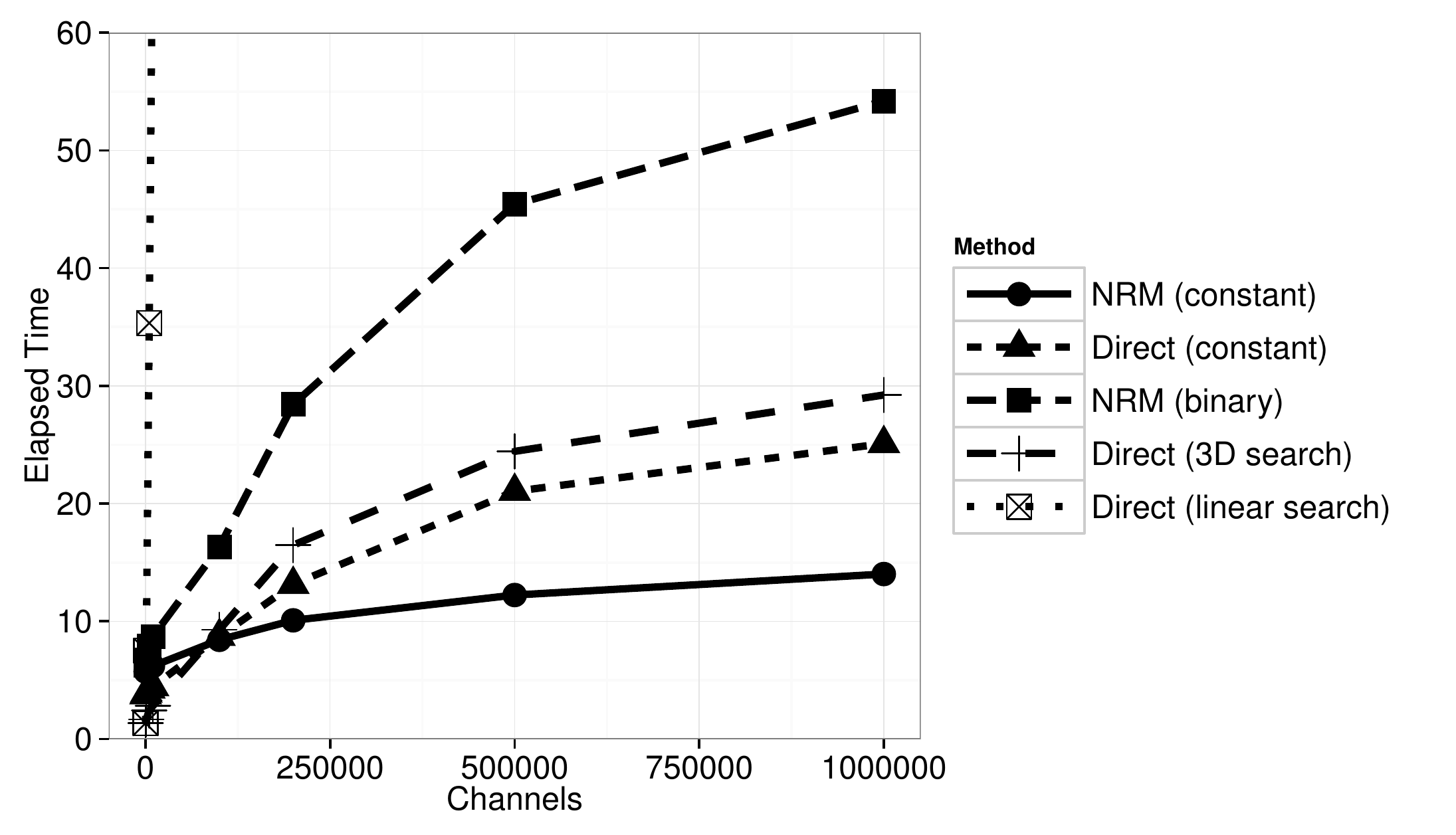}
\end{center}
\caption{Scaling on large problems. Under the same conditions as Figure \ref{fig:fig4} with larger problem sizes, the constant-complexity methods outperform the others. Although the O(1) algorithms scale roughly constant across a wide range of moderate problem sizes, as the problem size becomes large, the increased memory demands lead to imperfect scaling.}
\label{fig:fig5}
\end{figure}

As shown in Figure \ref{fig:fig4}, methods utilizing simple data structures with low overhead perform best on small to moderate sized problems. The example in Figure \ref{fig:fig4} used a random network model of unit-rate Poisson processes with a relatively high degree of connectivity (10 updates required for each step; note that the data structure updates were performed as if the propensities were changing, even though they were always set to unit rates.). The original NRM, implemented with a binary min-heap, would perform better relative to the others if fewer updates were required at each step. The constant-complexity NRM exhibits small timing fluctuations due to the method being tuned for much larger problems. In Figure \ref{fig:fig5}, we see that the constant-complexity direct method and constant-complexity NRM method outperform the others on large problems, with the constant NRM performing best. However, we see that the $O(1)$ scaling does not appear constant across large problem sizes. This is due to the effects of using progressively larger amounts of memory. Running the same experiments on a different system architecture could lead to differences in crossing points between methods, but the overall trends should be similar.

\subsection{3D Spatial Model}

The next subvolume method is a popular method for simulating spatial models. The NSM is different from the other methods considered here in that information about the spatial structure is built in to the algorithm. Therefore, it does not make sense to test the NSM reaction generator independent of a spatial model and full solver implementation.

Here we compare the NSM, the constant-complexity direct method, and the constant-complexity NRM on a model using a 3D geometry comprised of equal sized cubic subvolumes with periodic boundary conditions. The reactions and parameters are shown below \cite{elf2004spontaneous}.

\begin{subequations}
\begin{align*}
E_A & \xrightarrow{k_1} E_A + A\\
E_B & \xrightarrow{k_1} E_B + B\\
E_A + B & \xrightleftharpoons[k_d]{k_a} E_AB\\
E_AB + B & \xrightleftharpoons[k_d]{k_a} E_AB_2\\
E_B + A & \xrightleftharpoons[k_d]{k_a} E_BA\\
E_BA + A & \xrightleftharpoons[k_d]{k_a} E_BA_2\\
A & \xrightarrow{k_4} \emptyset\\
B & \xrightarrow{k_4} \emptyset.
\end{align*}
\end{subequations}
\begin{subequations}
\begin{align*}
k_1=150\;s^{-1}, \;k_a & = 46.2\;(\mu M)^{-1}s^{-1}, \\
k_d=3.82\;s^{-1}&, \;k_4=6\;s^{-1}\\
D = 10^{-8} & cm^2s^{-1}
\end{align*}
\end{subequations}
\begin{equation*}
[E_A](0) = [E_B](0) = 12.3 \; nM.
\end{equation*}

The diffusion constant $D$ is equal for all species. The rate constant for all diffusive transfer events is therefore $D/l^2$, where $l$ is the subvolume side length \cite{elf2004spontaneous}. We note that the first two reaction channels are a common motif used to model processes such as protein production for an activated gene or, in this case, enzyme-catalyzed product formation in the presence of excess substrate, where the rate constant $k_1$ implicitly accounts for the effectively constant substrate population.

\begin{figure}[h]
\begin{center}
\includegraphics[scale=0.4]{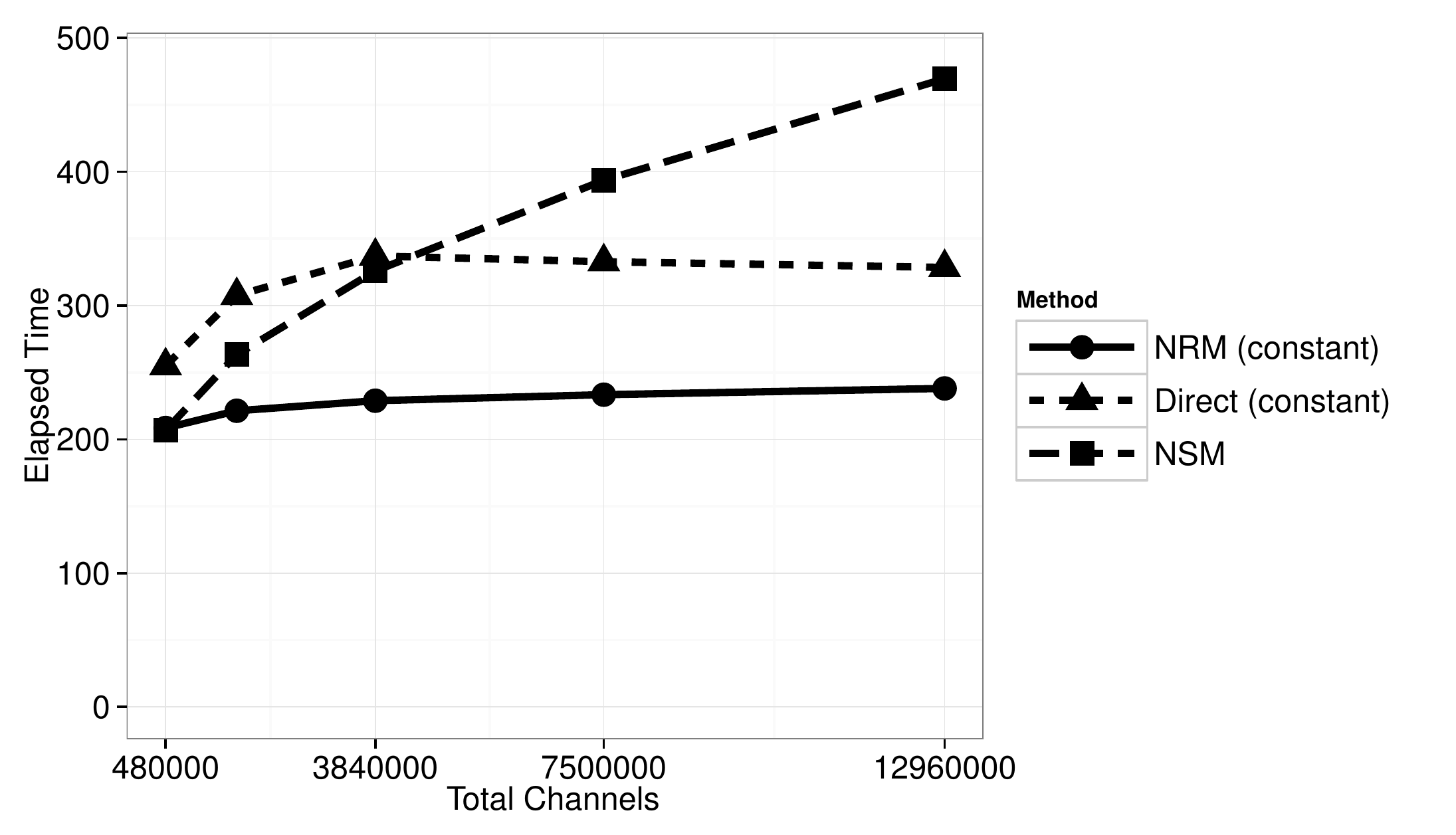}
\end{center}
\caption{3D spatial model with system volume $V = 12^3 \mu m^3$. Plotted is the elapsed time in seconds to execute $10^8$ simulation steps. Below $M = 480000$ channels, the NSM is more efficient than the constant-complexity methods. Above $M = 480000$ channels, the constant-complexity NRM is about 20\% faster than the constant-complexity direct method.}
\label{fig:fig6}
\end{figure}

It is possible to scale up the number of transition (reaction and diffusion) channels by changing the system volume or changing the subvolume side length. First, we consider a large volume, with domain side length $12 \mu m$ and subvolume side lengths ranging from $0.6 \mu m$ to $0.2 \mu$, corresponding to a range of $8000$ to $216000$ subvolumes, respectively (see Fig. N1C in the Supplementary Material of Elf and Ehrenberg \cite{elf2004spontaneous}). As shown in Figure \ref{fig:fig6}, the constant-complexity NRM outperforms the NSM and constant-complexity direct method for problems larger than $M=480000$, which corresponds to meshes finer than $20^3=8000$ subvolumes.

\begin{figure}[h]
\begin{center}
\includegraphics[scale=0.4]{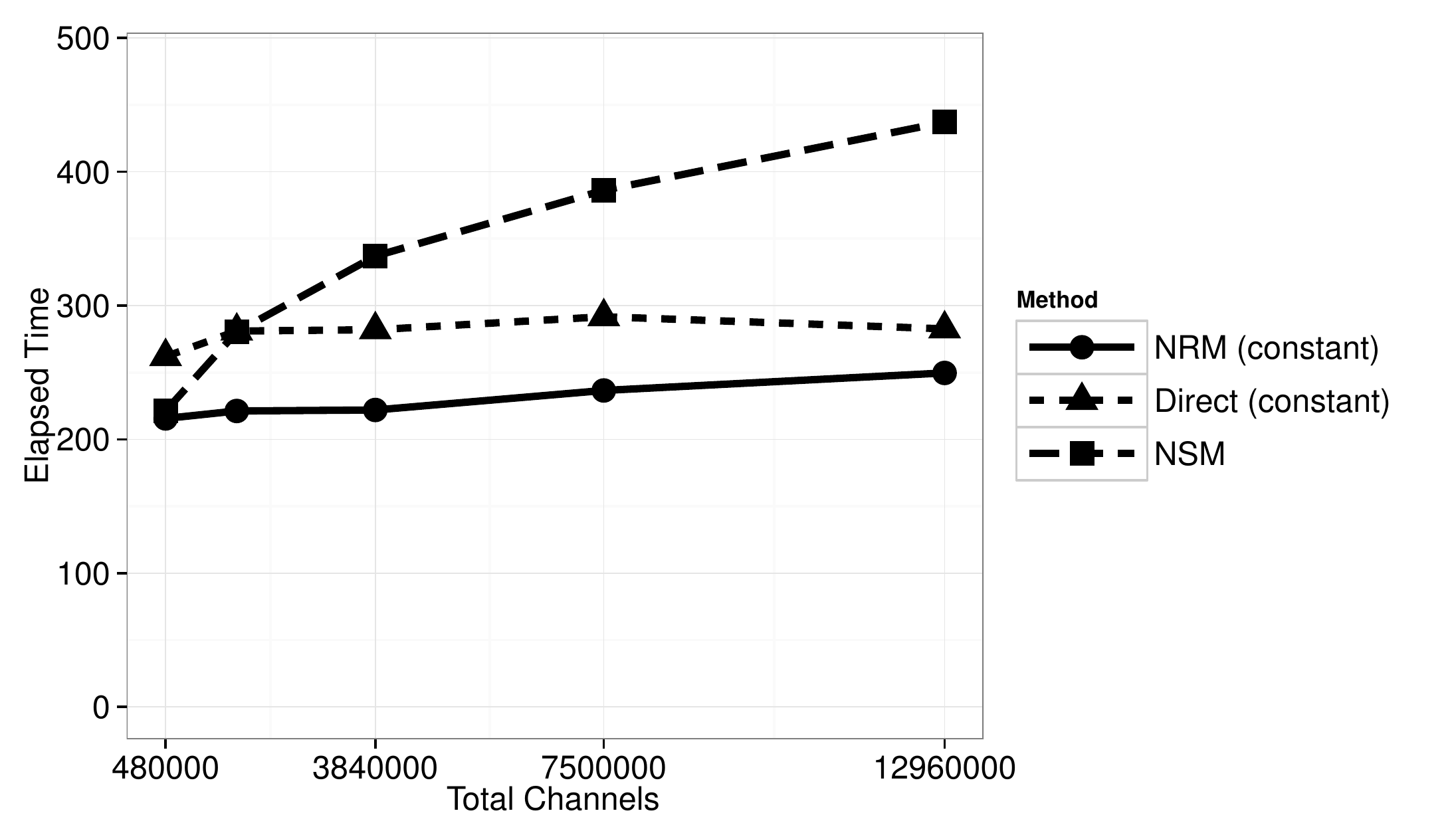}
\end{center}
\caption{3D spatial model with system volume $V = 6^3 \mu m^3$. At this smaller system volume, when the mesh is highly refined, most subvolumes are empty. Hence many of the reaction channels have zero propensity. The constant-complexity direct method and the NSM exclude all events with propensity zero from their data structures. The constant-complexity NRM uses the number of active (nonzero propensity) reactions to determine the number of bins $K$ to use, but this algorithm does not benefit much from having many zero-propensity reaction channels.}
\label{fig:fig7}
\end{figure}

We next consider the same 3D model with system volume $V = 6^3 \mu m^3$. As shown in Figure \ref{fig:fig7}, the constant-complexity NRM still achieves a benefit over the NRM and constant-complexity direct method, albeit a smaller improvement. In this example, there are approximately 28000 molecules in the system after the initial transient. At the finest resolution in Figure \ref{fig:fig7} there are 216000 subvolumes, of which most contain no molecules. Therefore, the majority of the reaction channels are effectively switched off, or inactive, with propensity zero. In this example at the finest resolution, typically fewer than $2\times10^5$ of the nearly $1.3\times10^7$ channels have nonzero propensities. The constant-complexity direct method and the NSM exclude zero propensity reactions from their reaction selection data structures, effectively reducing the problem size to the number of nonzero channels. The constant-complexity NRM does not benefit much from having many zero propensity channels.

Changing the spatial discretization influences the rates of $0^{th}$ order and bimolecular reactions and diffusion events. For instance, using a finer discretization increases the frequency of diffusion events. This means that more simulation steps are required to reach a fixed simulation end time. Changing the relative frequencies of different reaction channels influences simulation performance, though the effect is typically small for all methods. In the NSM, for example, increasing the relative frequency of diffusion events will improve performance slightly if there are fewer diffusion directions (e.g four for a 2D Cartesian mesh) than reaction channels for each subvolume because the average ``search depth" will be weighted more heavily toward the smaller diffusion event search.

\section{Discussion and Conclusions}
\label{sec:conclusions}

We have shown it is possible to formulate a constant-complexity version of the NRM that is efficient for exact simulation of large discrete stochastic models by using event time binning. Rather than targeting a load factor as one would with a hash table, the key to consistent efficiency of the algorithm is to choose the bin width based on the average propensity sum (and, therefore, step size) of the simulation. The examples in Section \ref{sec:numerical_experiments} demonstrate the advantages and some of the disadvantages of the constant-complexity NRM. The algorithm is not well suited for small models. However, for models with a large number of \textit{active} channels and timescales that do not vary too rapidly, the constant-complexity NRM is often more efficient than other popular methods.

For models with many inactive (zero propensity) channels, the performance of some SSA variants depends on the number of active channels rather than the total number of channels. The NSM scales proportional to the logarithm of the number of subvolumes containing active channels. The constant-complexity direct and NRM methods scale $O(1)$ in algorithmic complexity, but their performance does depend on the amount of memory used. Both constant-complexity methods have memory requirements for their reaction generator data structures that scale roughly proportional to the number of active channels. However, the table rebuild step in the constant-complexity NRM method scales as $O(M)$. This typically constitutes a small fraction of the total computational cost (e.g. $<3\%$ for the largest problem in Figure \ref{fig:fig5}). However, in the case of extremely large $M$ and an extremely small number of active channels, the relative cost of rebuilding the table in the constant-complexity NRM becomes more significant. In an extreme case, other methods such as the constant-complexity direct method, NSM, and original NRM may be more efficient.

It may be possible to modify the constant-complexity NRM to make it less sensitive to changes in the average simulation step size and number of active channels. The current dynamic table rebuilding strategy handles this well in many cases. However, in the case of extreme changes one could implement a ``trigger" that initiates a table rebuild if changes in the timescale or number of active channels exceeds a threshold. One could also envision utilizing step size data from previous realizations to guide the binning strategy, possibly even utilizing unequal bin sizes, to further improve performance.

Performance comparisons are inherently implementation, model, and system architecture dependent. While we have attempted to present a fair comparison and the general algorithm analysis is universal, the exact simulation elapsed times may vary in different applications. The constant-complexity NRM presented here is an efficient method in many situations but inappropriate in others. As modelers develop larger and more complex models and as spatial models become more common, this algorithm provides a valuable exact option among the large family of exact and approximate stochastic simulation algorithms. 

\begin{acknowledgments}
\noindent Research reported in this publication was supported by the National Institute Of General Medical Sciences of the National Institutes of Health under Award Number R01GM029123. The content is solely the responsibility of the authors and does not necessarily represent the official views of the National Institutes of Health.
\end{acknowledgments}

\appendix
\section{Algorithm Pseudocode}

The following pseudocode is representative of an implementation of this algorithm method.
\vspace{1em}

\noindent Model: propensities, $\nu$, dependencyGraph\\
\noindent DataStructure: table, lowerBound, binWidth, bins
\begin{algorithmic}[ht]
\Procedure{NRM}{x0, tFinal}
   \State $t \gets 0$
   \State $x \gets x0$
   \State $buildDataStructure()$
   \While{$t < tFinal$}
      \State $event \gets selectReaction()$
      \State $t \gets event.time$
      \State $x \gets x + \nu(event.index)$
      \State $updateDataStructure(event.index)$
      \State \% store output as desired
   \EndWhile
\EndProcedure

\vskip 1em
\Procedure{selectReaction}{}
   \State \% return min event time and index
   \State \% first, locate bin index of smallest event
   \While{$table(minBin).isEmpty()$}
      \State $minBin \gets minBin+1$
      \If{$minBin>bins$}
         \State $buildDataStructure()$
      \EndIf
   \EndWhile
   \State \% smallest event time is in table(minBin)
   \State \% find and return smallest event time and index
   \State \textbf{return} $min(table(minBin))$
\EndProcedure

\vskip 1em
\Procedure{buildDataStructure}{}
   \State $lowerBound \gets t$
   \State \% default 20*sqrt(ACTIVE channels)
   \State $bins \gets 20*sqrt(propensities.size)$
   \State \% default 16*step size
   \State \% in practice, an approximation to
   \State \% sum(propensities) is used
   \State $binWidth \gets 16/sum(propensities)$
   \For{$i =$1:$propensities.size$}
      \State $rate \gets propensities(i)$
      \State $r \gets exponential(rate)$
      \State $ eventTime(i) \gets t + r$
      \State $table.insert(i,eventTime(i))$
   \EndFor
   \State $minBin \gets 0$
\EndProcedure

\vskip 1em
\Procedure{updateDataStructure}{index}
   \For{$i$ in $dependencyGraph(index)$}
      \State $oldTime \gets eventTime(index)$
      \State $oldBin \gets ComputeBinIndex(oldTime)$
      \State $rate \gets propensities(i)$
      \State $r \gets exponential(rate)$
      \State $ eventTime(i) \gets t + r$
      \State $bin \gets ComputeBinIndex(eventTime(i))$ 
      \If{$bin \ne oldBin$}
         \State $table(oldBin).remove(i)$
         \State $table.insert(i,eventTime(i))$
      \EndIf
   \EndFor
\EndProcedure

\vskip 1em
\Procedure{table.insert}{i, time}
   \State $bin \gets computeBinIndex(time)$
   \State \% insert into array
   \State $table(bin).insert(i,time)$
\EndProcedure

\vskip 1em
\Procedure{ComputeBinIndex}{time}
   \State $offset \gets time - lowerBound$
   \State $range \gets lowerBound*binWidth*bins$
   \State $bin \gets integer(offset/range*bins)$
   \State \textbf{return} $bin$
\EndProcedure

\end{algorithmic}

%

\nocite{*}
\bibliographystyle{aipnum4-1}
\bibliography{sanftbib}

\begin{thebibliography}{38}
\bibitem [{foo()}]{footnote1}%
  \BibitemOpen
  \href@noop {} {}\bibinfo {note} {There is also a cost for the initial
  building of the data structure. This is at least O(M) for all methods as the
  propensity function for all reaction channels must be evaluated. However, the
  cost of building the data structure is relatively small for all algorithms
  for any simulation with a sufficiently large number of steps. There is also a
  computational cost to generate the random numbers. Generating uniform or
  exponential random numbers are $O(1)$ operations. The cost of generating
  random numbers can comprise a significant portion of the computational costs
  for simulating small models, but these $O(1)$ costs become insignificant for
  sufficiently large models where search and data structure update costs
  dominate.}\BibitemShut {Stop}%
\end{thebibliography}

\end{document}